# Effect of external electromagnetic radiation on the anomalous metallic behaviors in Ta thin films

Junghyun Shin, Sungyu Park[*] and Eunseong Kim[†]

Department of Physics, Korea Advanced Institute of Science and Technology, Daejeon, 34141, Republic of Korea.

We investigated transport characteristics of superconducting Ta thin films with three configurations in rf radiation filters; no filter, only room-temperature filters, and low-temperature filters in addition to room-temperature filters. The transport properties near the transition temperature are strongly dependent on whether the room-temperature filter is installed or not. The entire transition is shifted to higher temperature with loading layers of the room-temperature filters. Once the zero-resistance state is achieved at B=0, no strong radiation effect is observed even with low-temperature filters installed. When magnetic field is turned on, the nonzero-resistance saturation at low temperatures is revealed without low-temperature filters, which has been considered to be magnetic-field-induced quantum metallic phase. However, the insertion of the additional low-temperature filter weakens the saturation of the resistance, the signature of the metallic behavior. This observation suggests that the previously reported anomalous metallic state in Ta films is mainly induced by the unfiltered radiation and, thus, the intrinsic metallic ground state should be limited to the narrow range of magnetic fields near the critical point, if exists.

---

[*] Present address: Center for Artificial Low Dimensional Electronic Systems, Institute for Basic Science (IBS), Pohang 37673, Republic of Korea

[†] Corresponding author

## I. INTRODUCTION

The scaling theory of localization anticipates that the ground state of two-dimensional (2D) non-interacting electron systems is an insulating state because intrinsic disorders prohibit the appearance of a metallic state [1]. Strong correlation of electrons which allows the appearance of superconductivity can be observed marginally in a 2D superconductor because of its susceptibility to external perturbations such as disorders, currents, and magnetic fields ($B$'s). Accordingly, direct superconductor-insulator (SI) transition was theoretically expected and a metallic ground state was considered to exist only at the boundary between the two phases [2,3].

The SI transition has been investigated experimentally with various approaches. For instance, finite-size scaling analysis [4] near quantum critical point has been utilized to examine universal features of the SI transition [2,5]. Temperature ($T$) dependence of resistance at low $T$'s has been investigated to examine the ground state of disordered films [6]. Ac conductance [7], scanning tunneling microscopy [8], and Nernst effect measurements [9] have been conducted to identify the nature of the SI transition.

Nevertheless, an anomalous metallic phase at zero-$T$ limit has been observed. $B$-tuned metallic state was first observed in the study of amorphous MoGe thin films [10]. Under a wide range of $B$, decreasing resistance with thermally activated behavior at relatively high $T$'s transformed to $T$-independent finite-resistance saturation at low $T$'s. The metallic behavior has been widely reported in various thin films such as highly disordered InOx films [11], weakly disordered Ta films [12], and 2D crystalline films of 2H-NbSe$_2$ [13] and ion-gated ZrNCl [14].

The apparent metallic behavior at low $T$'s has been interpreted as appearance of Bose metal [15] and dissipation induced by quantum tunneling [16]. However, this interpretation has been challenged by the fact that unfiltered electromagnetic radiation can elevate electron $T$ which may disrupt the phase coherence in superconductivity. The skepticism has intensified by the recent observation of strong suppression of the metallic behavior by adequate filtering of radiation [17]. Installation of several layers of electromagnetic radiation filters eliminates remarkably the resistance saturation in both 2H-NbSe$_2$ and highly disordered InOx films. On the other hand, the more recent transport study on the YBCO thin films patterned with a triangular array of holes reported the existence of Cooper-pair phase coherence with finite-resistance saturation at low $T$'s with an adequately filtered measurement architecture [18,19]. Thus, the physical origin of the unexpected metallic behaviors has remained still controversial.

In this paper, we present radiation filtering effects on transport characteristics in amorphous Ta thin films. Superconducting Ta thin films are categorized to the family of weak-disorder samples with exceptionally smooth film morphology and show the intriguing metallic behavior at the lowest $T$'s among the films reporting the metallic state. The introduction of layered filtering architecture vastly reduces the dissipation below the transition $T$ ($T_C$) and, accordingly, metallic behavior. The phase diagram in the $B$-$T$ plane directly exhibits the suppression of the metallic phase which are intervening the superconducting and insulating phases.

## II. EXPERIMENTAL DETAILS

Ta thin films were deposited on $SiO_2$/Si substrates by using a DC magnetron sputtering method. The detail method of deposition and the characteristics from X-ray diffraction (XRD) and atomic force microscopy (AFM) measurements were reported previously [20]. Sputtered Ta thin films with thickness less than 5 nm show an amorphous structure and their spatial roughness is measured to be approximately 0.1 nm. The exceptionally low roughness generally attributes to the favorable wetting property of Ta thin films on $SiO_2$. The surface roughness of Ta thin films is comparable to 2D crystalline superconductors such as ion-gated ZrNCl [14], and much lower than highly disordered InOx [11]. Ta thin films reported in this study are listed in Table I. The three Ta films, Ta1-1, Ta1-2 and Ta1-3, were fabricated in the same batch and the Ta2-1 film was sputtered in another batch. The normal-state sheet resistance ($R_N$) for films deposited in the same batch has shown a gradual increase with decreasing film thickness. Small variation in $R_N$ from batch to batch has been observed despite the same nominal thickness.

TABLE I. List of superconducting Ta films. Film thickness ($d$) was monitored by a calibrated thickness monitor with a quartz microbalance. $R_N$ is $R$ near 10 K. $T_C$ is selected to $T$ when $R = 0.5\ R_N$. Despite of the same nominal thickness, Ta1-2 and Ta2-1 films show rather dissimilar $R_N$'s and, accordingly, different $T_C$'s. We found film morphology varies from batch to batch depending on details of fabrication conditions. Nevertheless, Ta films show consistent transport characteristics within the samples deposited during the same sputtering operation indicated as a batch number.

| Sample | Batch | $d$ (nm) | $R_N$ (kΩ) | $T_C$ (K) |
|---|---|---|---|---|
| Ta1-1 | 1 | 3.8 | 2.49 | 0.27 |
| Ta1-2 | 1 | 4.0 | 1.89 | 0.32 |
| Ta1-3 | 1 | 5.0 | 1.54 | 0.44 |
| Ta2-1 | 2 | 4.0 | 1.17 | 0.43 |

The Ta films were patterned into four-probe geometries for measurement by using a shadow mask during sputtering or electron-beam lithography with a subsequent dry-etching process after sputtering. The transport characteristic was measured by a standard lock-in technique with a measurement frequency of 7 ~ 15 Hz and an excitation current of 0.1 ~ 10 nA. A dilution refrigerator with a superconducting magnet was used to measure low-$T$ characteristics. $B$'s perpendicular to the sample plane were applied to study $B$ dependence. To investigate radiation filtering effects, three different configurations were used. (See Supplementary Fig. S1.) Initially, all samples were measured with a configuration without any low-pass filter (NF configuration). Then, room-$T$ filters (RTF's) were installed in a place right after the signal coming out of the fridge and before connected to the measurement lock-in amplifiers (RTF configuration). The RTF consisted with a home-made RC low-pass filter with a 100 kHz cut-off frequency and a commercial Pi filter with a 10 MHz cut-off frequency. Lastly, intensive filtration of additional low-$T$ filters (LTF's), RC low-pass filters, was set up on top of the RTF configuration on the mixing-chamber stage (AF configuration).

### III. RESULTS AND DISCUSSION

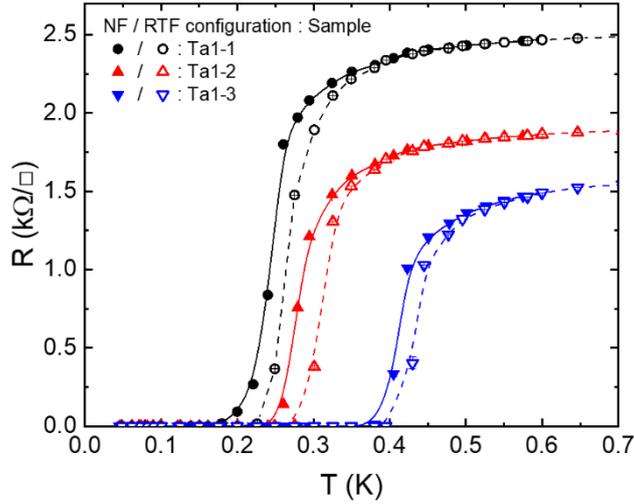

FIG. 1. $T$ dependence of $R$ at zero $B$ for Ta thin films, Ta1-1(black), Ta1-2(red), and Ta1-3(blue). Solid scatters show $R$(T) with the NF configuration. Open scatters are with the RTF configuration. The solid and dash curves are guides to the eye.

The low-$T$ transport characteristics were initially investigated without installing LTF's along cryogenic coax cables to examine the effect of RTF's. The measurement wires were properly heat-sunk at all stages of the dilution fridge. The $T$ dependence of the sheet resistance ($R$) of Ta thin films with (open symbols) and without (solid symbols) the RTF's at zero $B$ is shown in Fig.1. $T_C$ is determined when $R$ becomes 50% of $R_N$. $T_C$'s with the RTF's are 0.27, 0.32 and 0.44 K for Ta1-1, Ta1-2, and Ta1-3 respectively. $T_C$ decreases gradually with increasing $R_N$ as expected.

Elimination of the RTF's lowers $T_C$'s for the all three samples by approximately 25 mK compared to those measured with the RTF's loaded. This indicates that a typical measurement set-up without RTF's cannot prevent permeation of rf noise down to the sample stage successfully. The extensive effect by a single layer of the filter suggests the extreme susceptibility of the superconducting thin films to the external perturbations. Thus, this observation is essentially consistent with that reported in 2H-NbSe$_2$ where the removal of filters affects the entire $T$ dependence not only near $T_C$ but also at low $T$'s [17]. Pronounced difference in highly disordered amorphous InOx is that the installation of filters only influences the $T$ dependence of $R$ much below $T_C$ [17]. Marked discrepancy between InOx and 2H-NbSe$_2$ could be ascribed to the film morphology where the robust percolated superconducting islands in InOx films are less affected to the thermal fluctuations. In other words, the local $T_C$'s of superconducting islands exceed the average mean-field value of the thin film. Installation of the RTF's does not induce any noticeable change in $T$ dependence when superconducting Ta thin films exhibit zero-$R$ superconductivity at low $T$'s, suggesting fully developed superconducting thin films are immune (less susceptible) to the additional thermal fluctuation similar to the superconducting islands in InOx. Because no further effect on the transport properties at zero $B$ is observed, the importance of additional LTF's has not been emphasized in general [20].

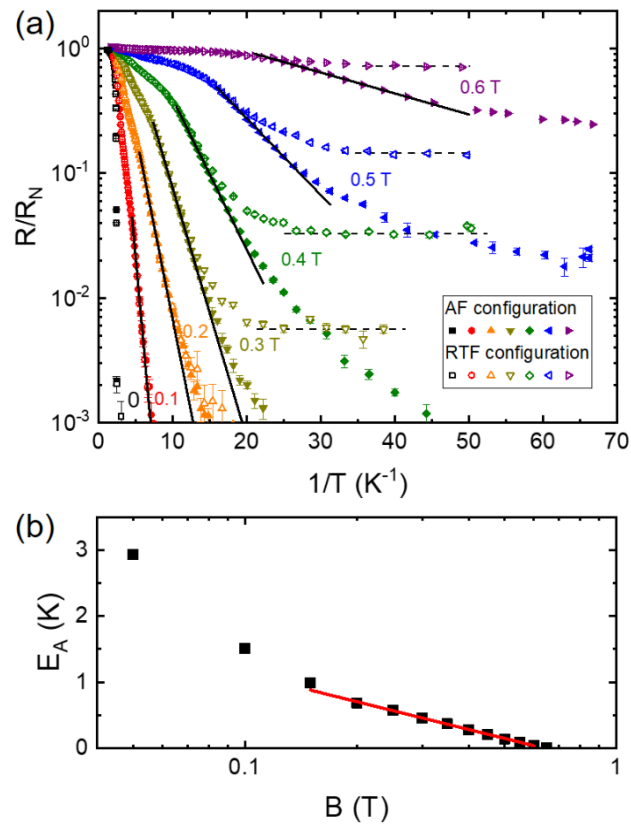

FIG. 2. $T$ dependence of $R/R_N$ under various $B$'s for T2-1. (a) Arrhenius plots under various $B$'s between 0 and 0.6 T with a 0.1 T step. Solid scatters are measured data with the AF configuration and open scatters are results with the RTF configuration. Black solid lines are results of linear fitting to $R/R_N \propto \exp(-E_A/T)$ and the slopes of the lines represent $E_A$'s. Black dash lines show the saturation of $R/R_N$ to finite values. (b) $B$ dependence of $E_A$ is shown in a semi-logarithmic plot. The red solid line is a linear fit to $E_A \propto \ln(B/B_0)$ where $B_0 = 0.64$ T.

Although the effect of LTF's is not distinct at zero $B$, it can be important when the $B$-induced resistive state appears. The $T$ dependence of the normalized resistance ($R/R_N$) of Ta2-1 under various finite $B$'s below the critical field ($B_C$) was investigated as shown in Fig. 2(a). The $T$ dependence of Ta2-1 with only the RTF's in an Arrhenius plot (open symbols) reproduced the intriguing intermediate metallic phase within a wide range of $B$'s. For instance, $R/R_N$ at $B = 0.4$ T exhibits a sharp drop at high $T$'s and the slope in an Arrhenius plot decreases gradually with saturation to a finite value at zero-$T$ limit, which has been regarded as the deviation from the conventional superconducting and insulating states.

Next, we investigated the effect of LTF's on the transport characteristics. The low-$T$ transport properties of the measurements with the LTF's (solid symbols) are directly compared with those without the additional LTF's. The $T$ dependence measured at $B$'s lower than 0.2T, which shows a relatively sharp decrease in $R/R_N$, demonstrates no apparent discrepancy between the two different configurations.

At $B \geq 0.3$ T, essentially identical $T$ dependence between the two configurations is observed at high $T$'s where both the configurations exhibit thermally activated behaviors. Additional filtering exerts a discernable influence at a low-$R$ region where $R$ drops to approximately 2 % of $R_N$ at $B = 0.3$ T. The intensively filtered film still follows thermally activated behavior (see, e.g., solid black lines) while the conventionally filtered film deviates from it. Further decreasing $T$ reveals that $R/R_N$ of the intensively filtered film can have two orders of magnitude smaller $R/R_N$ than that of the film without the LTF's at sufficiently low $T$'s. Besides, the installation of the LTF's destroys the finite-$R$ saturations shown in dashed lines which have been considered to be a hallmark of the quantum metallic phase. It is notable that $R/R_N$ of the thin film with the LTF's eventually deviates from the black solid line and, then, it further decreases to a lower value without obvious saturation. One may attempt to explain the deviation with the extreme susceptibility of the 2D superconducting films to external perturbations although the effect is now extremely minute. However, we cannot exclude the possibility of the quantum metallic state existing in the very narrow range of $B$'s near quantum critical point.

Thermally activated behavior at intermediate $T$'s can be characterized by a simple equation of $R/R_N \propto \exp(-E_A/T)$ where the activation energy $E_A$, a slope of the black line, is typically an average energy barrier for trapped vortices. Figure 2(b) shows $B$ dependence of $E_A$. The red solid line illustrates logarithmic dependence of $E_A$ on $B$, i.e., $E_A \propto \ln(B/B_0)$ where $B_0 = 0.64$ T. This logarithmic dependence is understood by the collective pinning model where the density of thermally activated dislocations in the vortex lattice mainly determines the activation barrier [10,21].

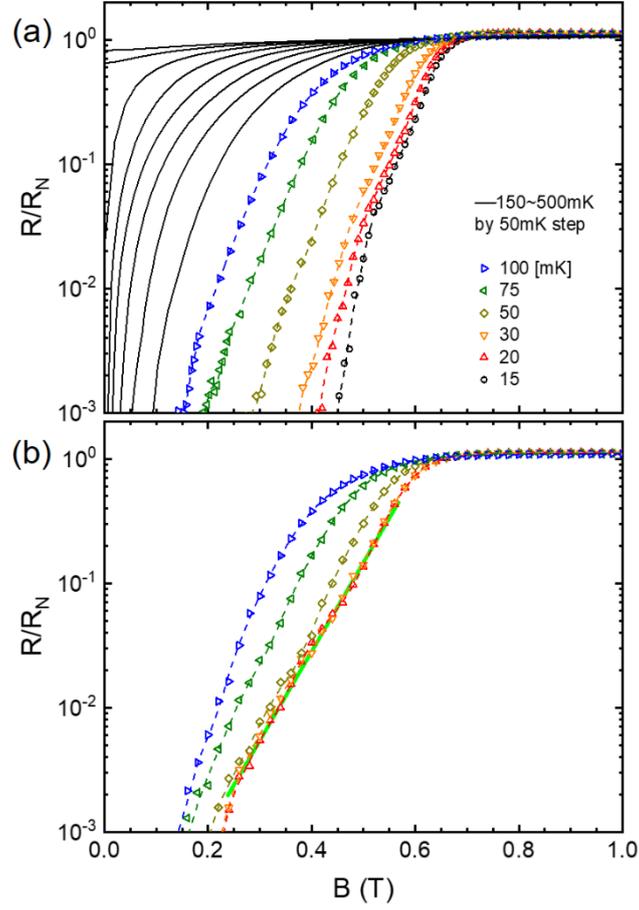

FIG. 3. MR isotherms at various $T$'s. (a) $R/R_N$ of Ta2-1 as a function of $B$ measured with the AF configuration. Black solid lines are measured at various $T$'s between 150 and 500 mK with a 50 mK step. Open scatters with different color codes are measured at $T = 100$ (▷), 75(◁), 50(◇), 30(▽), 20 (△) and 15 mK (○) respectively, and dash lines are guides to the eye. (b) $B$ dependence of $R/R_N$ with the RTF configuration. Open scatters are measured at $T = 100, 75, 50, 30$ and 20 mK, and dash lines are guides to the eye. The measurement $T$'s are displayed by the same color codes and symbols as those in (a). Green solid line exhibits exponential dependence of $R \sim \exp(B/B^*)$.

We also observed distinct filtering effects in magneto-resistance (MR) isotherm measurements. Figure 3(a) shows MR isotherm curves for Ta2-1 measured at various $T$'s with the AF configuration. Representative MR characteristics near the critical point, $B_C \sim 0.65$ T, where $dR/dT$ changes its sign were reproduced. The MR isotherm curves show steeper declines at lower $T$ with uniform $B$ dependence of which $R/R_N$ follows $\exp(B/B^*)$. One may recognize that the MR changes the slope at a certain $B$ where the dominant vortex dynamics may be altered with increasing vortex density. One may also speculate that the extreme susceptibility of 2D superconducting films at low $T$'s changes the $B$ dependence eventually.

The most striking observation is that the MR curves with the additional LTF's don't overlap with other isotherms even near the base $T$ below 20 mK, demonstrating that any saturated finite-$R$ at low $T$'s, the direct indication of the quantum metallic phase, is not present. On the contrary, MR isotherm curves less than 30 mK overlap well to a single curve in the measurements without the LTF's, which is consistent with the appearance of the saturated $R$ at low $T$'s with finite $B$ [see Fig. 2(a)]. The collapsed MR isotherms show single exponential field dependence of $R/R_N \propto \exp(B/B^*)$, as marked by a green solid line. In the MR isotherms measured at $T \geq 50$ mK, distinct separation from the collapsed isotherms is observed. For instance, the MR isotherm at $T = 50$ mK follows initially the collapsed isotherms and then branches off near 0.4 T. As further increasing $T$, isotherm curves at $T \geq 75$ mK are separated from the low-$T$ behavior at all $B$ ranges. The exponential $B$ dependence of saturated $R$ and the dissociation of high-$T$ MR isotherms are compatible with the previously reported MR measurements of amorphous MoGe [22].

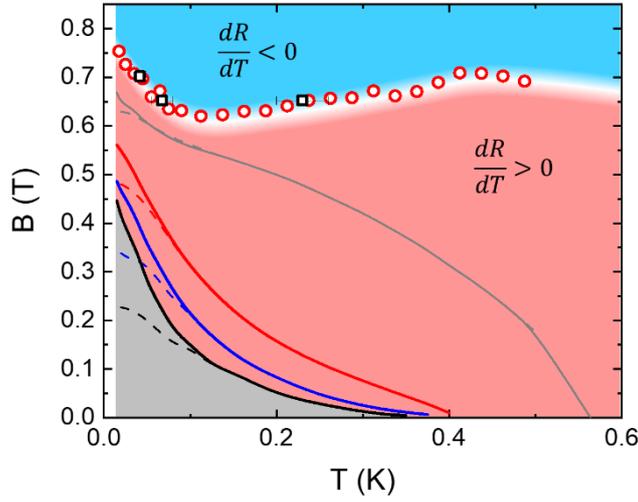

FIG. 4. Transport phase diagram of Ta2-1 in the $T$-$B$ plane. The diagram is constructed based on the measurements with the AF configuration, except dash lines. Red open circles, $B_{Cross}$'s, are crossing points of two adjacent MR isotherms [Fig. 3(a)]. Black open squares, $T_{Cross}$'s, are cross-over points between negative d$R$/d$T$ and positive d$R$/d$T$, obtained from $R(T)$ curves under 0.65 and 0.7 T (see Fig. S2 in Supplemental Material). A blue-shaded region above the open scatters shows negative d$R$/d$T$ while d$R$/d$T$ is positive in a red-shaded region. (See the main text for detailed discussions.)

The $T$ and $B$ dependences demonstrate that the $B$-induced metallic behavior in a Ta thin film can be vastly eliminated with intensive filtering. The crucial role of the LTF's can be marked additionally with examining the phase diagram of the Ta thin film in the $B$-$T$ plane. Figure 4 displays a reconstructed phase diagram based on the transport measurements with the AF configuration shown in Fig. 2(a) and Fig. 3(a). $B_{Cross}$ (red open circle) obtained from Fig. 3(a) is a crossing point between two adjacent MR isotherm curves, indicating that the Ta film at $B > B_{Cross}$ exhibits insulating at zero-$T$ limit. The boundary connecting $B_{Cross}$ points separates the insulating phase from the superconducting (or metallic) phase. Additionally, direct measurements of $R$ as a function of $T$ in the very narrow range of $B$ near $B_C$ reveal the crossing points, $T_{Cross}$'s, where the sign of d$R$/d$T$ is reversed. $T_{Cross}$'s are obtained by the $T$ dependence measurements at $B = 0.65$ T and 0.7 T (see Fig. S2 in Supplemental Material) and plotted in black open squares in the diagram. The cross points show excellent agreement with the SI phase boundary as shown in Fig. 4.

The SI transition at zero-$T$ limit can be understood within the framework of quantum melting of the vortex lattice. The phase boundary between superconductor and insulator near quantum critical point is expected to have relatively weak $T$ dependence. However, $B_{Cross}$ in the phase diagram exhibits steep increase with decreasing $T$ below 0.1 K, suggesting the destruction of the superconducting phase is more complicated by the interplay of quantum fluctuations, thermal fluctuations, and disorders in the thin film. The low-$T$ upturn of $B_{Cross}$ is remarkably similar to that observed in 3ML Ga films [23]. The presence of quenched disorders stabilizes the local vortex glasslike phases in this epitaxially-grown atomically-thin films. Accordingly, "global" superconductivity can be induced at extremely low $T$'s by long-range Josephson coupling of the superconducting islands. We speculate that the $T$ dependence of $B_{Cross}$ in the Ta thin film with relatively weak disorders can be understood within the essentially same framework of the vortex glasslike state at low $T$'s.

A blue-shaded area above the phase boundary determined by the red open circles and the black open squares shows a negative $dR/dT$ region while a red-shaded area indicates a positive $dR/dT$ region. The $T$ dependence of $R$ in the positive $dR/dT$ region is clarified further by iso-resistive curves of 0.9 $R_N$, 0.1 $R_N$, 0.01 $R_N$, and 0.001 $R_N$ which are shown by grey, red, blue, and black solid lines respectively. Dashed lines with the same color codes represent the transport measurements with only the RTF's with the corresponding values of $R$ to compare with the intensively filtered results. The all solid lines show stiff increases consistently below about 0.1 K. Similarly, the zero-$R$ region in our measurement limit shaded with grey color increases sharply, nearly identical to the metal-insulator boundary. On the other hand, the measurement with only the RTF's (the dashed lines) shows $R$ saturation because of thermal broadening at low $T$'s, which leads to the illusive appearance of a metallic phase. At the base $T$ of our dilution fridge, the zero-$R$ region is defined below ~0.45 T and extrapolated to 0.55 T at zero-$T$ limit. Therefore, the intrinsic metallic ground state, if exists, is expected to be only in a narrow range of $B$'s near $B_C$.

The reduction of metallic phase with intensive filtration indicates that Ta thin films can be as sensitive as 2D crystalline 2H-NbSe2 to external perturbations. Ta thin films with thickness less than 5 nm are microscopically disordered such as amorphous InOx films. However, the surface roughness in Ta films is order of magnitude smaller that of InOx films. We speculate that the vortices are pinned rather weakly in a Ta thin film due to this surface morphology with low vortex pinning barriers. Weakly pinned vortices can be agitated by small perturbations easily and moved with resistive dissipation, which leads to the broadening of the transition with finite-$R$ saturation even below 0.1 K. It seems consistent with the extreme sensitivity of the vortex state in amorphous MoGe films to external rf radiation [24]. Appropriate filtration of electromagnetic radiation results in dramatic suppression of the mobility of vortices. In addition, because electron-phonon scattering in 2D superconducting films near zero $T$ becomes extremely weak, presumably thin films could be extremely susceptive to even small external radiation. Although Ta films are disordered, the exceedingly low $T_C$ compared to that of typical InOx films may allow to measure the marked difference.

However, the reduction of thermal fluctuations with extensive filtration is not sufficient enough to eliminate metallic phase completely. The deviation from thermally activated dependence was entirely eliminated with the installation of LTF's in Ref. 17. Nevertheless, the low-$T$ deviation is still present in the Ta thin film. Non-Arrhenius $T$ dependence of $\exp[-(E_0/T)^p]$ with $p < 1$ (see short-dot lines in Supplementary Fig. S3) was obtained in a disordered InOx film [11], which could be ascribed to variable-range-hopping resistivity induced by quantum vortex tunneling [25]. Besides, continuous change of a dominant thermally activated mechanism with decreasing $T$ is another possible explanation to understand the deviation [21]. Dissipation in a vortex lattice arises due to the phase fluctuations mediated by the displacements of dislocations, which has two $T$-dependent terms; the rate of thermally activated dislocation-creep over pinning barriers and the density of dislocations which consist of thermally activated dislocations as well as disorder-induced dislocations. In the weak collective pinning region where the interaction energy of vortices is more significant than the characteristic pinning barrier of disorders, $R$ mainly depends on the density of thermally activated dislocations at sufficiently high $T$'s, which can explain the logarithmic $B$ dependence of $E_A$ as shown in Fig. 2(b). At low $T$ when only disorder-induced dislocations remain, the main $T$-dependence term is the thermal creep rate of surviving dislocations, which will appear as thermally activated dependence with another characteristic activation energy from pinning barriers.

In summary, we studied the effect of radiation on the metallic phase in superconducting Ta films with the three different filtering configurations (NF, RTF, and AF). Significantly reduced metallic phase is obtained in the transport measurements with the AF configuration, demonstrating that the unfiltered thermally agitated fluctuations are mainly responsible to the elusive "metallic" phase. Since thermally activated process with decreasing $T$ was not fully recovered with intensive filtering, there could be other possible explanations for the low-$T$ transport properties in Ta thin films. To clarify the $T$ dependence at low $T$'s, it is necessary to investigate electron $T$ with different filtering configurations.

**ACKNOWLEDGMENTS**

This work was supported by the National Research Foundation of Korea (NRF) Grant funded by the Korean Government (MSIP) (NRF-2016R1A5A1008184) and the MSIT(Ministry of Science and ICT), Korea, under the ITRC(Information Technology Research Center) support program(IITP-2020-2018-0-01402) supervised by the IITP(Institute for Information & Communications Technology Planning & Evaluation).

# Supplemental material

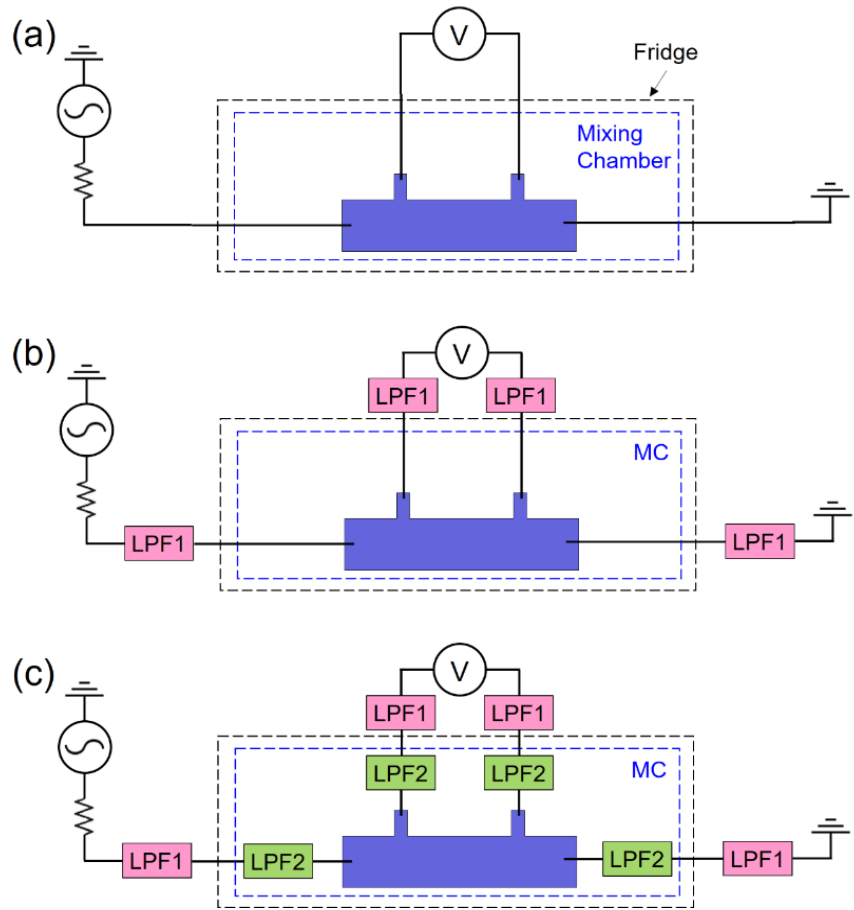

FIG. S1. Schematic diagrams for three different measurement configurations. Purple-colored shape with 4-wire connection represents the Ta thin film. The sample mounted at the mixing chamber (MC) in the fridge is electrically connected to a lock-in amplifier for transport measurement. (a) NF configuration. The measurement configuration without any filters. (b) RTF configuration. The measurement was performed with only room-temperature (room-$T$) filters. (c) AF configuration. Both room-$T$ and additional low-$T$ filters are installed. LPF1 represents a room-$T$ filter installed between the measurement equipment and the top of the fridge, which consists of a RC low-pass filter with a 100kHz cut-off frequency and a Pi low-pass filter with a 10MHz cut-off frequency. LPF2 is a RC low-pass filter installed on the MC.

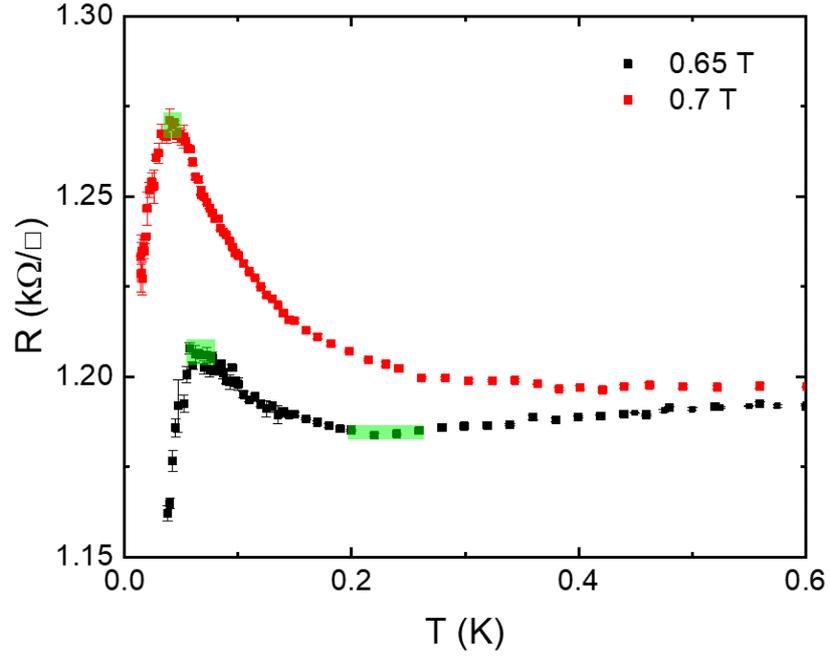

FIG. S2. $T$ dependence of sheet resistance ($R$) under magnetic fields of 0.65 T and 0.7 T with the AF configuration. Square green boxes, $T_{Cross}$, exhibits the crossing-over region between positive d$R$/d$T$ and negative d$R$/d$T$, which is represented in Fig.4 of the main text as black open squares.

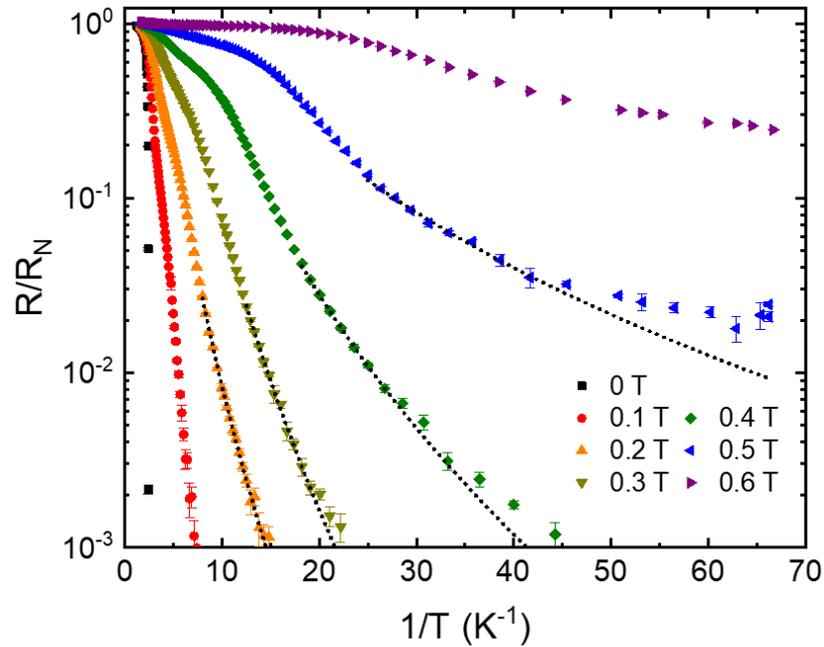

FIG. S3. Arrhenius plots for $T$ dependence of normalized resistance ($R/R_N$) under various magnetic fields (0~0.6 T), where $R_N$ is the normal-state sheet resistance. Solid scatters are the normalized resistance with the AF configuration, the same data to those in Fig. 2(a). Black short-dot lines are fitting curves for $R \propto \exp(-(T_0/T)^{1/3})$, where $T_0$ is a constant in units of Kelvin.